# Topological model for complex shape parts machining

*Abstract: Complex shapes are widely used to design products in several industries such as aeronautics, automotive and domestic appliances. Several variations of their curvatures and orientations generate difficulties during their manufacturing or the machining of dies used in moulding, injection and forging. Analysis of several parts highlights two levels of difficulties between three types of shapes: prismatic parts with simple geometrical shapes, aeronautic structure parts composed of several shallow pockets and forging dies composed of several deep cavities which often contain protrusions. This paper mainly concerns High Speed Machining (HSM) of these dies which represent the highest complexity level because of the shapes' geometry and their topology. Five axes HSM is generally required for such complex shaped parts but 3 axes machining can be sufficient for dies. Evolutions in HSM CAM software and machine tools lead to an important increase in time for machining preparation. Analysis stages of the CAD model particularly induce this time increase which is required for a wise choice of cutting tools and machining strategies. Assistance modules for prismatic parts machining features identification in CAD models are widely implemented in CAM software. In spite of the last CAM evolutions, these kinds of CAM modules are undeveloped for aeronautical structure parts and forging dies. Development of new CAM modules for the extraction of relevant machining areas as well as the definition of the topological relations between these areas must make it possible for the machining assistant to reduce the machining preparation time. In this paper, a model developed for the description of complex shape parts topology is presented. It is based on machining areas extracted for the construction of geometrical features starting from CAD models of the parts. As topology is described in order to assist machining assistant during machining process generation, the difficulties associated with tasks he carried out are analyzed at first. The topological model presented after is based on the basic geometrical features extracted. Topological relations which represent the framework of the model are defined between the basic geometrical features which are gathered afterwards in macro-features. Approach used for the identification of these macro-features is also presented in this paper. Detailed application on the construction of the topological model of forging dies is presented in the last part of the paper.*

## 1. Introduction

The various parts developed for aerospace, automotive and electrical or electronic appliances have complex shapes. These shapes can be designed using curved surfaces or portions of canonical surfaces whose arrangement is complex (different directions followed by several intersections). Shapes complexity associated with curvature variations generate difficulties during parts manufacturing or dies and molds machining used in casting, injection and forging. Dies and molds generally reproduce the complexity of the parts shape and machining difficulties that result. When analyzing the machining process, two complexity levels can be highlighted (Fig. 1). The first complexity level is between quasi prismatic parts and the aeronautic structures parts. Pockets (or cavities) generally shallow which characterize these aeronautic structures parts are made up of flat bottoms limited by thin walls whose orientations vary sometimes with respect to bottoms normal vectors [2]. The second complexity level leads to forging dies which are characterized by rather deep cavities including sometimes protrusions. This paper mainly concerns High Speed Machining (HSM) of these dies which represent the highest complexity level because of the shapes' geometry and their topology [3].

Fig. 1   Complexity levels of machined shapes

Machining of the complex shape parts is carried out on 5 axes machine tools ensuring a good continuity of the orientation of the cutting tool [4] [5]. For the shapes found on the forging dies, a 3 axes machine can be sufficient to ensure the machining. High Speed Machining (HSM) allows in this case improving productivity (reduction in machining time) and surface quality. But it induces an important preparation time, particularly during the analysis stages of the CAD model. Indeed, tasks carried out at these stages for a wise choice of cutting tools and machining strategies are time consuming.

Current machining process of complex parts consists of five types of machining operations: roughing, semi-finishing, finishing, re-machining and a prospective polishing [6]. Roughing aims to closely approach the final shape of the part by ensuring a maximum material rate and smooth transitions between tool paths. Geometries associated with this type of operation are pockets or cavities defined by levels. Semi-finishing must result in a constant machining allowance or a constant engagement of cutting tool during the next operation. Finishing operation is calculated on dies final geometry while integrating the respect of specified quality. Machining recovery is associated to transition shapes which generally have low curvature and small areas. It is performed by cutting tools with small diameters [7]. Implementation of the HSM process therefore requires the development of a complex process planning that is very different from that presented in most of the work on prismatic parts machining [8] [9]. Indeed a process planning for prismatic parts defines the sequence of simple machining operations, each including the geometrical feature and machining parameters related to machine tool, cutting tool and cutting conditions [10]. Sequence of operations is based on features accessibility constructed from the adjacency graph describing parts topology [11]. In the case of dies, it is difficult to identify geometrical features which are the base of the process planning. Several works have been made in the extraction of geometric characteristics of complex shapes. They are generally oriented design support of parts [12] [13], determination of cutting tool orientation to improve cutting conditions 14] and tools paths generation without planning or scheduling [15] [16]. In the case this paper is focused, complex shapes processed correspond to those found on the forging dies. Works results in this field allow identifying machined surfaces, cutting tools used, but with difficulty machining limits. At this stage, basic machining operations can be generated but it is difficult to ensure that tools paths are gouge-free and to optimize trajectories, in particular those out of material. Adjacency graph of surfaces and shapes associated with an evaluation of their sizes and positions is expected to solve the difficulties. But the concavity test used to build adjacency graph for prismatic parts is not applicable to complex shapes (large variations in curvature and normal direction). It is therefore appropriate to see how topological information required to generate process planning of complex shapes can be extracted or defined from CAD model.

In this paper, a model developed for the description of the topology of complex shape parts is presented. It is based on machining areas extracted for the construction of geometrical features starting from CAD model of the parts. As topology is described in order to assist machining assistant during machining process generation, the difficulties associated with tasks he carried out are analyzed at first. The topological model presented after is based on the basic geometrical features extracted. Topological relations which represent the framework of the model are defined between the basic geometrical features which are gathered afterwards in macro-features. Approach used for the identification of these macro-features is also presented in this paper. Detailed application on the construction of the topological model of forging dies is presented in the last part of the paper.

## 2. Machining process generation

Approach for generating a machining process of complex shape parts is widely based on a set of tasks, some of which are carried out using modules available in the CAM software. These tasks can be gathered in three main processing modules highlighted in the generic structure of the machining process (Fig. 2): a geometrical pre-processor, a geometrical processor and a geometrical post-processor. Description and exploitation of topological information are made throughout the development of the different stages of the machining process generation. Parameters taken into account both in the description and exploitation of the topological information are presented in the following.

Fig. 2    Structure of the machining process generation

### 2.1    Geometrical pre-processor

Geometrical pre-processor has several objectives. First, it allows characterizing machining areas with regard to the part geometry, kinematics of a given machine tool and the parameters related to cutting tools while integrating machining difficulties which can be encountered. On the other hand, topological relations between different machining areas should be defined in this geometrical pre-processor. Available CAM software provides few tools and modules to automatically extract machining areas and define the topological relations. Therefore, the machining assistant mainly relies on its own know-how to perform these tasks.

#### 2.1.1    Extraction of machining areas

Extraction of machining areas is carried out after the topological decomposition of the part CAD model (in STL format). At this level, considered topology is the orientation of part shapes compared to the cutting tool direction and the trajectory that can be followed during machining. Formalization of problems associated to this type of topology reflects the difficulties often encountered by machining assistant [17]. Aggregation of machining areas obtained by bringing closer maps corresponding to decomposition levels (cutting tool orientation and continuity in machining feed direction) make it possible to define three geometrical features types:

- Flank: geometrical feature that has a relatively small angle with the tool axis (machining direction) and which surface is quasi vertical during machining.

- Bottom: geometrical feature that has a wide angle with the tool axis and which surface is quasi

- Transition: geometrical feature that has a variable angle with the tool axis. It provides in general a link between flanks and bottoms as blend surfaces with fixed or variable radius.

Geometrical features identified according to these three types correspond to machining areas with their characteristics corresponding to machining difficulties integrated in the identification process.

Fig. 3   Examples of geometrical features

The following geometrical features were extracted in the example shown in (Fig. 3): four bottom features, three flanks and five transition features. These geometrical features are identified at several levels of the part and on convex and concave shapes. Their graphic visualization with "performance viewer" tool already presented in our previous work [18], makes it possible to assist machining assistant during the other tasks of the machining process generation. Tasks carried out in the pre-processor finish with the definition of topological relations between geometrical features in order to describe their relative positions. This description is an important step because results of the following tasks of the machining process generation will depend on its fineness and completeness.

Before introducing the description of the topological relations, it is necessary to highlight their impact on tasks like cutting tools selection, machining strategies choice, tool paths computation and their planning.

Fig. 4   Description of the relative position of shapes

*2.1.2   Impacts and scope of topological data*

Machining process generation of complex shape parts using any CAM software begins with the selection of the machined surface and shapes to be avoided. Machined surface must be understand as the set of geometrical features considered for which it is sometimes necessary to define machining limits (Fig. 4a). Two set types of geometrical features can be defined: cavities corresponding to the concave shapes and protrusions composed of convex shapes (Fig. 4b). Roughing is usually carried out by considering cavities (material removal rate) up to the protrusions (obstacles to avoid). During semi-finishing and finishing, cavities and protrusions are the basic shapes that are grouped together during machining when selected cutting tools and machining strategies make it possible. When these basic shapes cannot be grouped, they are selected as machining limits or obstacles (protrusion).

Topology description corresponds to different machining problems linked to interference check for the set cutting tool/tool holder. Two types of interference are considered: local interferences that occur when active parts of cutting tools fit into the material (Fig. 5.a.) and global interferences which are associated to contacts between tool holder and machined cavities or protrusions located in the cavities (Fig. 5.b.). Local interferences are processed during tool paths computation and planning based on machined surfaces topology [19] [20] [21]. Global interferences are not often processed by tool paths computation algorithms [22] [23]. Machining simulation or machining assistant expertise is often needed to process this type of interference [24]. Topology description proposed in this paper will be useful for the identification of machined areas or shapes involved in this type of interference.

Fig. 5   Cutting tool/part interferences: a. local, b. global

Generally in all CAM software, tool paths computation is initially carried out through a selection of the machined surface and the identification or selection of avoidance shapes. When any avoidance shape is selected, tool paths are limited to the boundaries of the machined surface. In the case of a cavity linked to the machined surface (topological relation), tool paths can be extended in order to maintain the set point of the feed rate (Fig. 6.a). However, extension of tool paths over the cavity generates input/output material that can lead to the damage of surface roughness. Unlike the cavity, a protrusion belonging or linked to the machined surface becomes an obstacle which requires a tool paths deviation (Fig. 6.b). Identification of protrusions is not automatic in CAM software. The definition of topological relations provides useful information to automation of gouge-free tool paths generation. Currently, protrusions selection as avoidance shapes is made by the machining assistant during topological analysis of part geometry. Identification of a protrusion as an avoidance shape is even more difficult when it is not directly linked to the machined surface (Fig. 6.c). In this case, only the extension of adjacency relations between geometrical features associated with the estimation of the height of the protrusion make it possible to define the avoidance area.

Fig. 6   Impact of the relative position of shapes on tool paths

Machining assistant usually selects machining strategies which limit feed rate reductions in order to ensure the geometric quality of the machined part. Example shown in Fig. 7.b presents tool paths simulations during machining of a cavity composed of two flanks and one bottom feature. Two types of machining strategies adopted are analyzed: the classic parallel plane strategy and the guided curve strategy using the two curves limiting the bottom feature by the two flanks (Fig. 7.a). As shown in these simulations, the influence of the topology relations between geometrical features induces more or less feed rate reductions that are not located in the same areas of the part. For the parallel plan strategy, feed rate reductions associated with several inputs/outputs material are located near the flanks while the middle of the tool paths is relatively spared. Conversely, the guided curve strategy generates feed rate reductions in the middle of the tool paths while sparing flanks. The location of feed rate reductions shows that they are associated to the extension of low curvature effects induced by the flanks (Fig. 7.b). Unlike the visual perception that give results on both types of strategy, analysis of the classification of tool paths lengths shows that the guided curve strategy generates longer tool paths and thus limits the number of inputs/outputs material (Fig. 7.c). This classification allows comparing the benefits of feed rate reductions and effects on increasing the number of inputs/outputs. When cavities are open, the tool paths can be extended in order to remove (or refer below) feed rate reductions at the beginning and the end of the tool paths (Fig. 7.d). This open or closed cavity concept is important for topology relations' description.

Fig. 7   Topology impact on feed rate reduction

Influence of the topology on feed rate reduction and increase in the number of input/output material is highlighted in the examples shown in Fig. 7. However, it seems difficult to solve all the machining difficulties towards a simple description of topology relations. This description can provide invaluable assistance to the selection of machining strategies and the identification of components required for tool paths computation (curved guides that support the algorithm for tool paths computation).

In the proposed approach for machining process generation, topological relations are defined between geometrical features obtained by topological decomposing of CAD geometric model based on maps of tools /parts contact areas. Geometrical features and topological relations between them represent final results of the geometrical pre-processor which are also basic data for cutting tools and machining strategies selection used for machining process generation. When these resources are consistently selected [17], topological relations are used to validate cutting tool dimensions (length and diameter) and machining strategies refinement (identification of guide curves, defining extension areas of tool paths ...). Results of the pre-processor module are very important because they correspond to the creation of machining features which must be processed independently (see Fig. 2).

*2.2   Geometrical Processor and post-processor*

The geometrical processor is used for computing tool paths and building machining sequences based on information provided by the geometrical pre-processor. Tool paths generation can be considered as fully automated when machined surfaces (cavities and protrusions), avoidance surfaces (protrusions) and resources (cutting tools/tool holders and machining strategies) are known. Machining and cutting conditions are basic data defined by the machining assistant. Machine tools and machining configuration types (3 or 5 axes) are the machining conditions determined according to parts shapes. Cutting speeds and feed rates (cutting conditions), are computed according to parts material and the couple material/cutting tool.

Tool paths computation which is the most automated task in CAM software resulted in machining cycles, each of which is associated to a cavity or a protrusion. Tool path planning which follows is aimed to gather machining cycles associated with a single cutting tool in order to create machining sequences. Any change in a machining sequence that is not linked to the cutting tool leads to the creation of a sub-sequence. Thus, when a cavity and a protrusion integrated in the same machining sequence are machined with two different strategies, each of them will be associated with a strategy sub-sequence.

When machining sequences are defined, creation of numerical control (NC) program is mainly carried out by converting control points on tool paths in NC codes. This is a conversion task performed by the geometrical post-processor which integrates tool calls, cutting conditions (cutting speed and feed rate) and specific instructions associated to the machine tool and the Numerical Control Director (NCD).

**3.   Topological model**

*3.1   Basic elements of the topological model*

Geometrical features created during the topological decomposition of a given forging die are the main elements of the topological description. They are gathered in macro-features such as cavities and protrusions which are the basic elements of the topological model. A cavity or a protrusion is composed of one flank and one or more bottoms features (Fig. 8). Their difference lies in the concavity/convexity property that characterizes the geometrical features gathered. Concavity/convexity condition is defined using the Gaussian curvature (K) and medium curvature (H). This approach has been developed in several works carried out on the identification of geometrical features [25] [26] and its application leads to the definition of six types of shape (Fig. 8a). Analysis of the different geometrical features makes it possible to highlight two cases:

- Geometrical features which satisfy a uniform concave or convex condition. Their shape properties are set to respectively concave (Cv) or convex (Cx).
- Geometrical features with several concave and convex or flat areas. As the shape property is unspecified in this case, only the analysis of the geometrical features gathered will indicate whether a cavity or a protrusion.

Analysis of Gaussian and mean curvatures of the part presented in Fig. 8b makes it possible to gather the geometrical features in three macro-features: one cavity, two protrusions and the parting surface corresponding to an isolated bottom feature. In the case of some parts more complex than the example, identification of macro-features can be difficult when several cavities share the same transition feature. A proposed approach for identifying macro-features is presented in the following.

Fig. 8 Principle of topology description

*3.2 Topological relations*

According to the basic elements of the topological model, the topological structure of a given part has two levels: geometrical features (level 1) gathered in macro-features (level 2). Topological relations are consequently defined at each of these two levels. They are detailed in the following towards a formal description.

*3.2.1 Topological relations between geometrical features*

Considering only the basic geometrical features, the obvious topological relation which can be defined is the adjacency relation representing contact between the geometrical features. Indeed, the environment of a geometrical feature is very local, since in most cases it does not allow itself to create a cavity or a protrusion. The Gaussian and mean curvatures property of the geometrical features nevertheless makes it possible to identify the adjacency relation. This identification is very useful in the extraction process of macro-features and as shown above can guide the selection of cutting tools and machining strategies.

Adjacency relation between two geometrical features is characterized by sharing a common edge. Beyond this common edge, it is important to determine, if possible, the concavity or convexity property of the adjacency relation through the transition feature which often provides the link between bottom features and flanks. Starting from the transition feature positioned between a flank and a bottom feature, three types of adjacency relations can be highlighted (Fig. 9). Concave adjacency relation based on a closed transition feature (Fig. 9.a) leads to the construction of a cavity. Conversely, the convex adjacency relation also based on a closed transition feature (Fig. 9.b) is used to create a protrusion. Unspecified adjacency relation shown in Fig. 9.c is based on a transition feature which seems concave at first sight. But curvatures analysis of this transition feature shows that it corresponds rather to a saddle valley lying on a bottom feature with variable curvatures.

Fig. 9 Adjacency topological relations

A closed transition feature that depends on the closed edge it shares with a flank or bottom feature is useful information. Indeed, this closed transition feature isolates the cavity or the protrusion from the local topology allowing its direct identification even starting from flanks or bottoms with an unspecified topological property. UML graphs associated with different topological adjacency relations will be further exploited in the extraction of macro-features. Properties of flank and bottom features are not used for the characterization of adjacency relations and the extraction of macro-features. They will be hidden in the next UML graphs for easier reading.

*3.2.2 Topological relations between macro-features*

Description of the topology of macro-features is intended to represent at first their relative positions, which as shown above can have an influence on tool paths deviation and collision detection. Then, it is important to set a propagation path of the topological relations to other macro-features that are not linked directly to the given macro-feature. Analysis of several parts and machining difficulties associated to the topology relations presented above makes it possible to highlight two types of relations: the relative position of two macro-features through their basic geometrical features (flank and bottom feature) and the configuration of protrusions inside cavities.

The first relation defining relative position "superposition" is created when a macro-feature which can be a cavity or a protrusion lean on a same another macro-feature through a bottom feature (Fig. 10.a) belonging to one of the two macro-features. The bearing surface can sometimes be spanning the bottom feature and the flank but in all cases, the edge which limits the transition feature and the bottom feature and/or the flank is a closed curve. This topological relation reflects a stack of macro-feature that can lead for example in important depths (cavities) or heights (protrusions) that would not be significant considering macro-features separately. The second relation defining relative position "open onto" has almost the same characteristics as the first relation (Fig. 10.b). The difference lies in the bearing surface which is generally a flank and the edge which limits the transition feature and the bottom feature and/or the flank is not closed. When it is defined between cavities, the topological relation "open onto" describes a crossing of one of the cavities in the other allowing tool paths extension. If a protrusion "opens onto" another protrusion, both define a super-protrusion which can be processed like only one protrusion.

Fig. 10   Macro-features topological relations

The two configuration relations define heights ratio of a cavity and a protrusion which is located in the cavity through an adjacency relation. In the case of the relation "belongs to" the height of the protrusion $H_{pr}$ is less than that of the cavity $H_{cv}$ (Fig. 10.c). Machining of any geometrical feature located outside the cavity having the relation can be carried out without taking into account the protrusion isolated in the cavity. On the other hand during the machining of the cavity, this relation will require determining cutting tool dimensions (diameter for example) according to the size and the relative position of the protrusion. The last configuration relation "oversteps" is characterized by a height of the protrusion $H_{pr}$ higher than that of the cavity $H_{cv}$ (Fig. 10.d). In this case, dimensions and the relative position of the protrusion must be taken into account not only when determining dimensions of the cutting tool selected for the machining of the cavity, but also for all the cavities located above the first when part of the protrusion arrived in their field. This situation is the typical case of propagation of the first relation "oversteps" towards the other cavities which are linked to the first through superposition relations. In all examples shown in Fig. 10, initial unspecified property of transition features was transformed after analysis of the local topology. Approach used for automatically transforming adjacency relation properties of these transition features is presented further (see §3.3.1).

Using UML formalism provides a framework for a more formal description of the topological relations defined in this work. Some components of the topological relation model such as exclusion conditions are defined by data encapsulated in the UML modelling module. They are not represented on the figures. For a given part, definition of the topological relations makes it possible to build its topological relation graph. Approach used for this construction is presented in the following.

*3.3   Topological relations graph*

Topological relations graph is aimed to provide a topological representation of a complex shape part relevant to machining assistant when generating the machining process of the part. This graph can also be integrated into CAM software to provide partial automation of tool paths computation and planning of machining sequences. Approach used for building the topological relations graph has the two following steps. In the first, the construction of the primary relations graph between geometrical features gives basic elements and structure for the identification of macro-features. In the second step, the final graph is generated by extension of the primary relations graph. This is done by integrating topological relations between macro-features that have been previously identified.

*3.3.1   Construction of the primary graph*

Construction of the primary graph is classical since it corresponds to the identification of geometrical features sharing a common edge. Nodes of this graph represent geometrical features and common edges of these geometrical features are designed by arcs (Fig. 11.a). As we have stated previously, the definition of adjacency relations is based on transition features. Adjacency relation is created when a transition feature has common edges with several other geometrical features which can be flanks or bottom features. General case of this adjacency relation corresponds to two geometric features such as flanks or bottom features sharing common edges with a single transition feature. When more than two geometrical features such as flanks or bottom features are linked to one transition feature, the adjacency relation is complex and it must be processed using a specific approach. The property (concave or convex) of the adjacency relation must then be specified on the graph. When the adjacency relation is constructed based on a transition feature which is concave or convex, it takes this property (respectively concave or convex). Otherwise (unspecified transition feature), the property of the adjacency relation should be determined based on the relative orientation between the geometrical features such as flank or bottom features.

Fig. 11   Primary graph of topological relations

Identification of the adjacency relations' properties defined from an unspecified transition feature is carried out automatically based on analysis of the material angle. Analysis plane V is created from any test point $P_1$ of the common edge of the transition feature and one of the two geometrical features such as flank or bottom feature defined in the adjacency relation. This plane is parallel to machining direction and perpendicular to the projection in the plane H of the tangent vector to the common edge at point $P_1$ (Fig. 11.b). Intersection of the plane V with the second common edge determines the second test point $P_2$ (the nearest point from $P_1$ when intersection results in several points). $n_1$ and $n_2$ are the normal vectors of the two geometrical features such as flank or bottom feature from the points $P_1$ and $P_2$. The last two test points $Q_1$ and $Q_2$ correspond to the minimum distance points between two lines $L_1$ and $L_2$, created respectively from the couples ($P_1$, $n_1$) and ($P_2$, $n_2$). The two test points $Q_1$ and $Q_2$ can be the same intersection point of the two lines $L_1$ and $L_2$ when it is possible. Property of the adjacency relation is finally determined from the following conditions:

If $\overrightarrow{P_1 Q_1} \cdot \vec{n_1} < 0$ and $\overrightarrow{P_2 Q_2} \cdot \vec{n_2} < 0$ then adjacency relation is convex (Fig. 11.b).

If $\overrightarrow{P_1 Q_1} \cdot \vec{n_1} > 0$ and $\overrightarrow{P_2 Q_2} \cdot \vec{n_2} > 0$ then adjacency relation is concave (Fig. 11.c).

Only these two conditions are used for geometrical features extracted from CAD models of forging dies because of their specificity. For more complex parts, the points $Q_1$ and $Q_2$ can not be determined when for example a transition feature is adjacent to two flat bottom features that are parallel. In this case the set of the three geometrical features will be considered a same machining feature which integrates the topology information for machining process generation. Generally, distances ($P_1$, $Q_1$) and ($P_2$, $Q_2$) reflect the relative orientation between geometrical features such as flank or bottom features linked to a given transition feature. Indeed, when these distances are important, orientation angle between the geometrical features will be low (quasi parallel surfaces). Machining assistant may set a distance threshold to merge the geometrical features.

### 3.3.2 Construction of the final graph

Identification of macro-features is carried out when starting the construction of the final graph before topological relations are defined. This identification is carried out in four steps. The final graph is designed at the fifth step of the process.

Fig. 12 Principle of macro-features identification

***Step 1:*** *Analysis of adjacency relations with the part's bounding box*

Geometrical features that have a common edge with the bounding box of the part are identified at this stage and their adjacency relations with other geometrical features are temporarily hidden. In the case of forging dies or similar shapes, these geometrical features are gathered in the parting surface which in fact has a "superposition" topological relation with the cavities of the die main cavity to be identified. An adjacency relation not necessarily associated to a transition feature is automatically created between the bounding box of the part and the adjacent geometrical features. In the example shown in Fig. 12, bottom 1 is identified at this step and the adjacency relation associated with transition 1 is temporarily hidden.

***Step 2:*** *Iterative identification of simple cavities and protrusions*

At the first iteration, bottom geometrical features which have only one adjacency relation are identified. Gathering the identified bottom feature with a flank geometrical feature and the transition feature associated to the adjacency relation makes it possible to create a protrusion or a cavity according to the property of the transition feature. When cavities and protrusions are created, all their external adjacency relations are temporarily hidden and a new iteration is started. In the example shown in Fig. 12, protrusion 1 and protrusion 2 are identified in the first iteration and adjacency relations associated with transition 3 and transition 4 are hidden. In the second iteration, cavity 1 is created.

***Step 3:*** *Identification of cavities and protrusions with multiple flanks*

A protrusion or cavity is generally composed of a single flank. When a part has shallow areas with variable curvature, transitions features can be inserted between different portions of flanks leading to protrusion or cavities with multiple flanks. In a given primary graph or a graph obtained by decomposition of complex shapes, this type of macro-feature with multiple flanks corresponds to a bottom feature which has several adjacency relations with flanks. Each adjacency relation is associated with a single transition feature. Protrusions and cavities with multiple flanks are identified after simple macro-features.

***Step 4:*** *Processing of complex macro-features*

When a given adjacency relation is shared by several geometrical features such as flank or bottom features through the same transition feature, it results in a complex macro-feature that should be decomposed. The decomposition performed by the machining assistant involves separating first bottom features or flanks associated to more than two adjacency relations. When the geometrical feature is separated, each geometrical sub-feature is associated to at most two adjacency relations. Transition features which are associated to more than one adjacency relation are then decomposed into sub-transition features each one associated to a single adjacency relation. Adjacency relations are updated to integrate results from the decomposition of complex macro-features. Example of decomposition is carried out in the industrial application presented in the last part of the paper. When all complex macro-features are decomposed, the processing goes back to step 2 for the identification of new simple macro-features.

*Step 5: Construction of the final graph*

When all protrusions and cavities are identified, adjacency relations temporarily hidden during the identification process are restored. Topological relations between identified macro-features are defined exclusively from the hidden relations that have just been restored. The definition of the topological relations is performed semi-automatically. First, topological relations presented in §3.2.2 are defined based on the analysis of the macro-features linked and the characteristics of each relation. Next, machining assistant can validate or modify the proposed topological relations because they are relevant for the machining process generation. In the example shown in Fig. 12, the following topological relations are defined: protrusion 1 "oversteps" cavity 1 and protrusion 2 "belongs to" cavity 1. These topological relations are complemented by one that reflects the fact that the parting surface represented by bottom 1 "is superposed to" cavity 1. This last relation generates a propagation of the "oversteps" relation. According to this propagation, protrusion 1 automatically "belongs" to the parting surface. Heights evaluation makes it possible to transform this relation: protrusion 1 "oversteps" the parting surface. In the final graph (Fig. 12), basic geometrical features and their adjacency relations are embedded in the identified macro-features for easier reading.

The topological relations graph created for a given part represents its topological model used in CAM software to generate machining process. Complete automation of the construction of this graph is possible, but requires access to geometrical data of the CAD model which is not easy for complex shapes. In the current version of the implementation system, construction of the graph is disconnected from the CAD modeller used (CATIA V5). However, the primary graph is designed automatically from data extracted from the CAD modeller. The implementation system assists the machining assistant in the definition of the topological relations.

## 4. Application

In this section, the proposed approach is used to build the topological model of a forging die used in industry to produce steering arms for vehicle. Geometrical decomposition of the forging die CAD model which was carried out in previous works leads to the following sixteen geometrical features: six bottom features, four flanks and six transition features (Fig. 13). Complexity of the shape of the forging die is not related to the basic surfaces (canonical surfaces) but to geometrical features orientations, relative positions and depth variations.

Fig. 13 Identification of simple and complex macro-features

Analysis of geometric features extracted from machining assistant point of view shows that flank 1 and transition features connected with it have a complex topology in that they are designed from bottom 4 to bottom 2 through bottom 3. This complexity is also clearly visible on the adjacency graph through the lot of topological relations involving flank 1 and the transition 2 (Fig. 13). Some transition features having an unspecified shape property, the construction of the primary topological graph involved analysis of the material angle. This primary graph shows five simple adjacency relations with two concave adjacencies and three convex adjacencies. Adjacency topological relations cannot be defined in this stage from transition 2 because of its complexity.

Identification of macro-features begins in step 1 by associating bottom 1 to the bounding box of the forging die. This bottom feature which represents the parting surface is superposed to cavities that will be extracted from the die main cavity. Step 1 ends with the creation of the convex adjacency relation between flank 1 and bottom 1, which is immediately hidden. When starting step 2, the geometrical feature that has only one adjacency relation is bottom 6. Gathering bottom 6, flank 4 and concave transition 6 makes it possible to create cavity 1 during the first iteration. The convex adjacency relation associated with the transition 5 is hidden at the end of this first iteration. Bottom 5 is the geometrical feature which has only one adjacency relation at the beginning of the second iteration. Cavity 2 is constructed by gathering bottom 5, flank 3 and the concave transition 4. The convex adjacency relation associated with transition 3 is hidden at the end of the second iteration. When starting the third iteration, there is no bottom feature with one adjacency relation. As any macro-feature with multiple flanks can be identified in step 3, the set {flank 1, flank 2, bottom 2, bottom 3, bottom 4, transition 2} represents a complex macro-feature that requires specific processing in step 4.

Fig. 14   Decomposition of a complex macro-feature

Step 4 starts with the decomposition of the complex macro-feature. According to this decomposition, flank 1 is divided into three geometrical sub-features {flank 1.1, flank 1.2, flank 1.3} (Fig. 14). This first decomposition is performed by machining assistant in the fields of bottom 2, bottom 3 and bottom 4 as shown in Fig. 13. Information on potential areas for the decomposition is given through analysis of the primary graph. This information is provided for machining assistant in the form of queries. After the decomposition of flank 1, transition 2 which has multiple adjacency relations with several geometrical features is also divided into five geometrical sub-features {transition 2.1, transition 2.2, transition 2.3, transition 2.4, transition 2.5} (Fig. 14). The processing goes back to step 2 for the identification of cavity 3, cavity 4 and cavity 5 with multiple flanks (flank 1.3 and flank 2).

In step 5, adjacency relations associated with transitions 1, 2.4, 2.6, 3 and 5 which were temporarily hidden are restored. They are processed in order to define topological relations between macro-features identified previously. The final topological graph of the forging die (Fig. 15) highlights on the one hand, a superposition of cavities 1, 2 and 3 which is "opened onto" cavity 4. This cavity is also "opened onto" cavity 5. Cavities 3 and 4 are "opened onto" each other because there is no flank between them. The topological model of the forging die thus created (Fig. 15) provides valuable assistance to the machining assistant.

Fig. 15   Final topological graph of the forging die

## 5.   Conclusion

A topological relations model used for describing configurations and relative positions of geometrical features is presented in this article. These geometrical features are extracted from the CAD model of complex shape parts such as forging dies. Description of topological relations is carried out to assist machining assistant during the generation of machining process. Assistance is particularly relevant because the topological relations were defined from the analysis of the difficulties encountered by the machining assistant. The first level of the topological relations is generated from the adjacencies between geometrical features. Adjacency relations created at this level represent geometrical features that sharing common edges. Properties of these relations which can be concave or convex are defined from the shape properties of transition features defined between the basic geometrical features. When a transition feature has an unspecified shape property, a method is proposed for analyzing the relative orientation of the basic geometrical features associated to the transition. Results of this analysis change the unspecified property into concave or convex adjacency relation. The other topological relations are defined between macro-features which gathered geometrical features in order to create cavities and protrusions processed in CAM software. A semi-automatic approach for identifying these macro-features is also proposed. It is based on the adjacency relations in the automatic stage but involves machining assistant to decompose flanks and transition features which sometimes belong to several macro-features. Analysis of the topological graph provides information on the areas where the decomposition must be done. Topological relations between macro-features are defined from the analysis of their relative positions and their characteristic dimensions. As analysis data cannot be automatically extract directly from CAD model of the part, these relations are defined by the machining assistant. This work is based on queries generated from an automatic pre-processing of the graph under construction. The final topological graph of the part created mostly after several iterations is the expression of the topological model of the part.

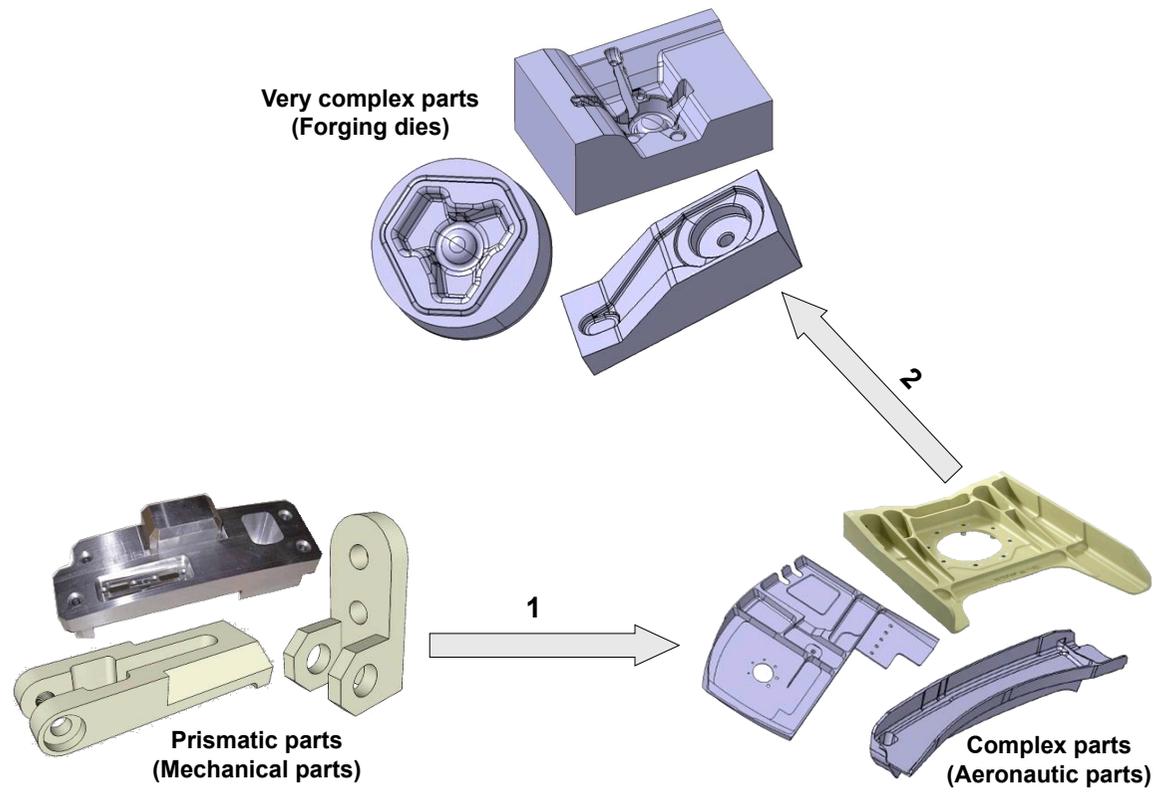

**Fig. 1 Complexity level of the machined shapes**

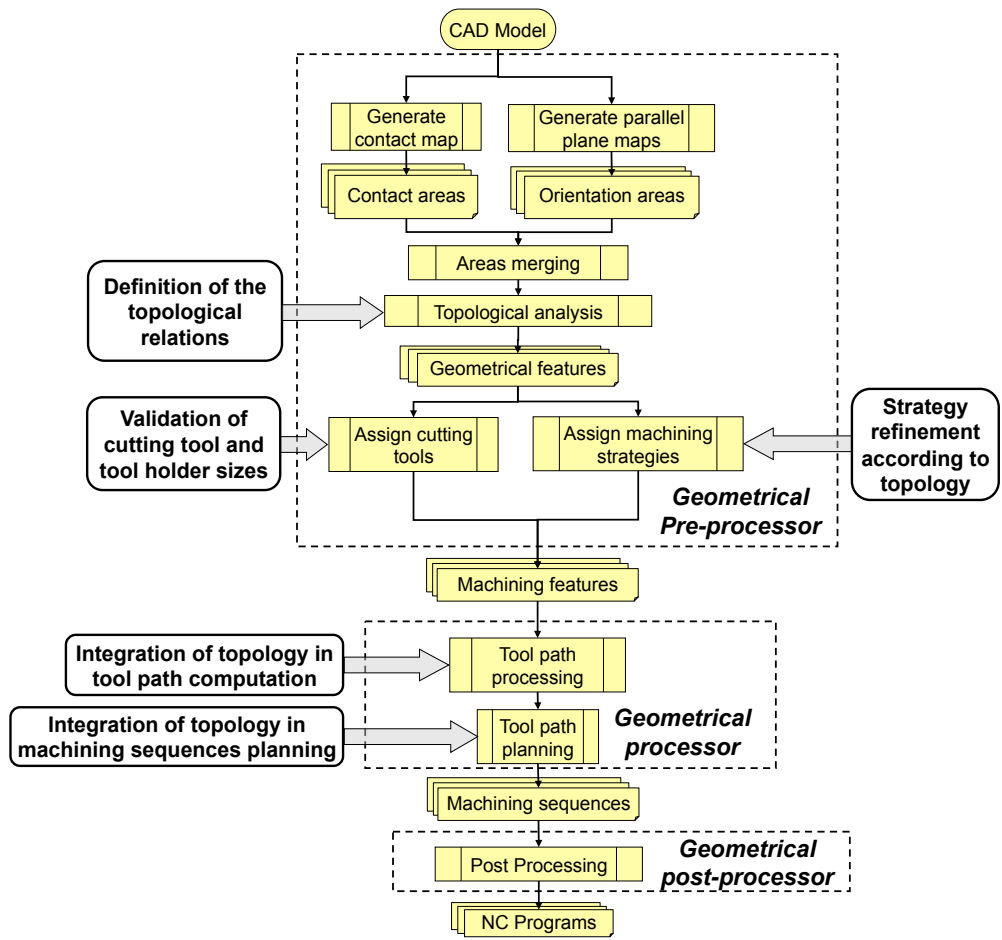

**Fig. 2 Machining process generation**

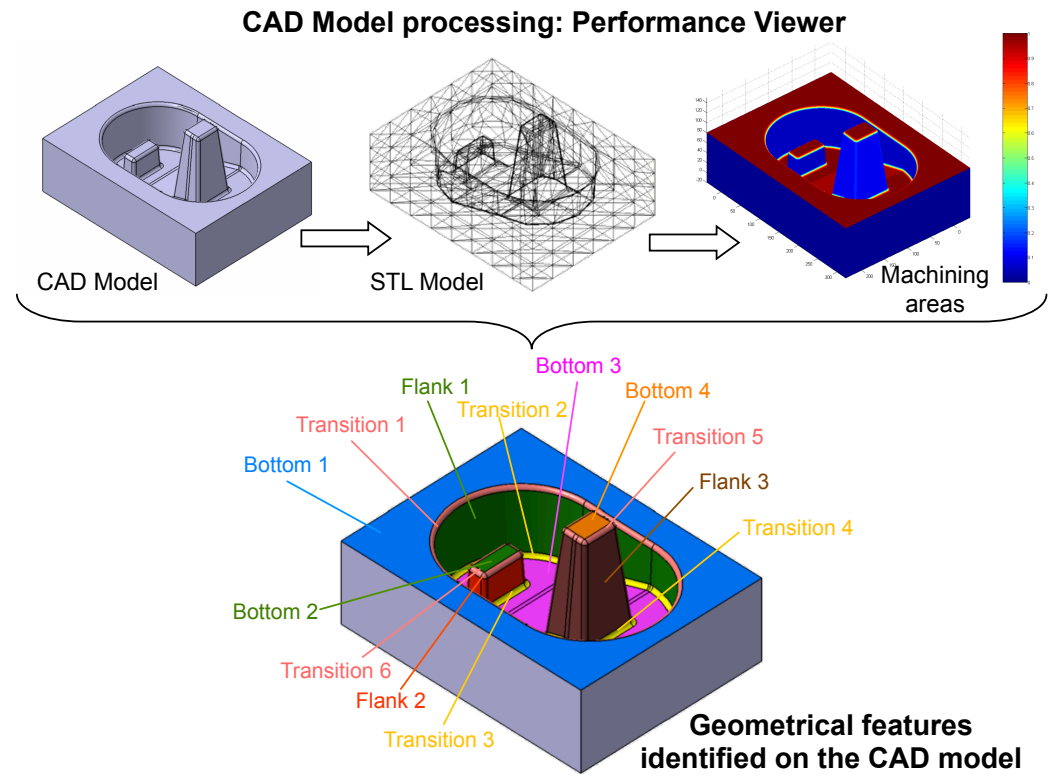

Fig. 3 Geometrical features identification

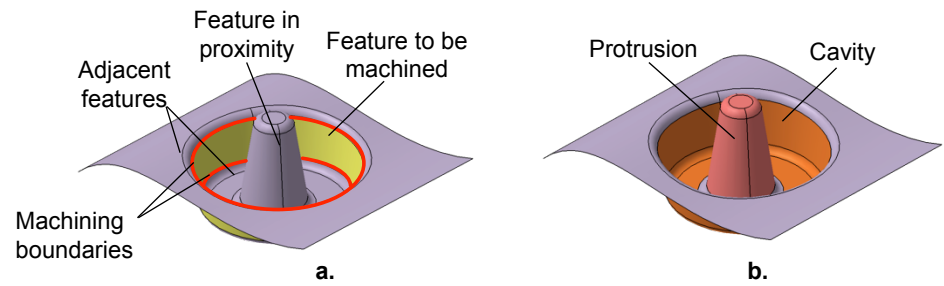

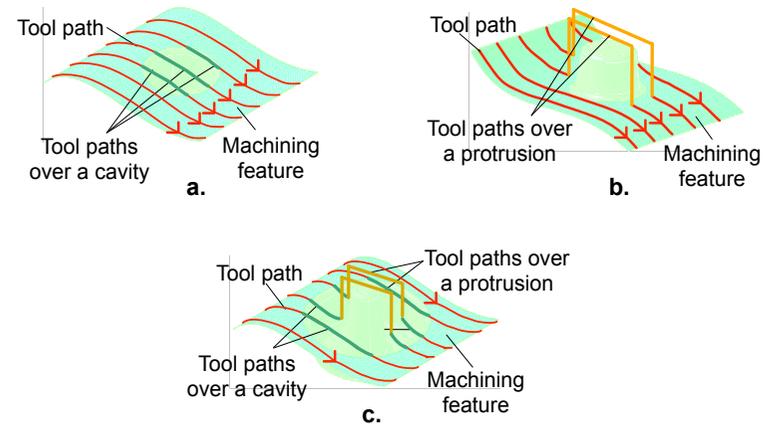

Fig. 4 Shapes and proximity of geometrical features

Fig. 6 Influence of proximity on machining tool path

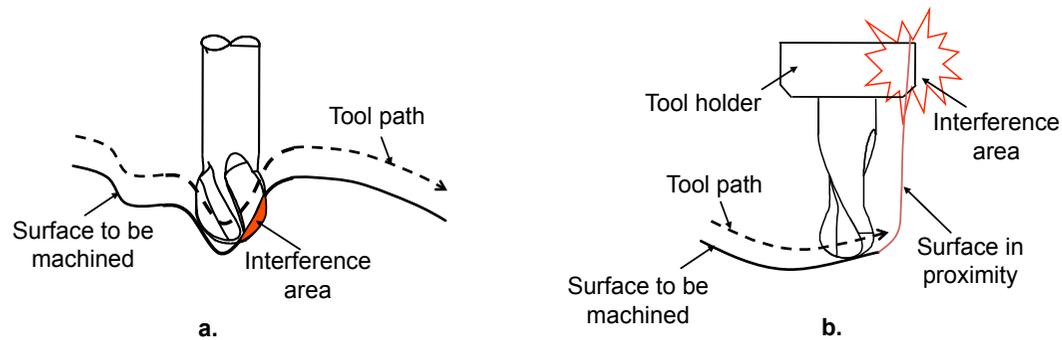

Fig. 5 Interferences tool/part

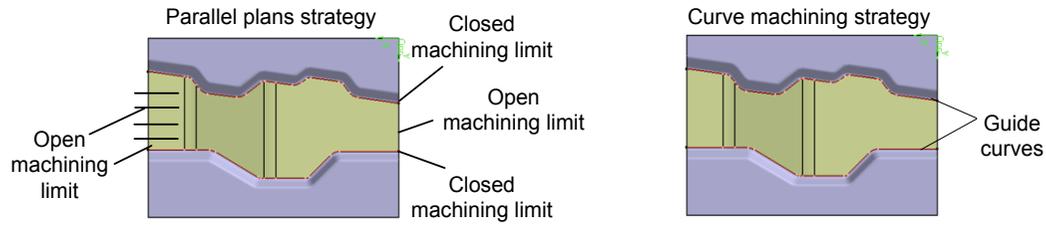

**a. Machining strategy choice**

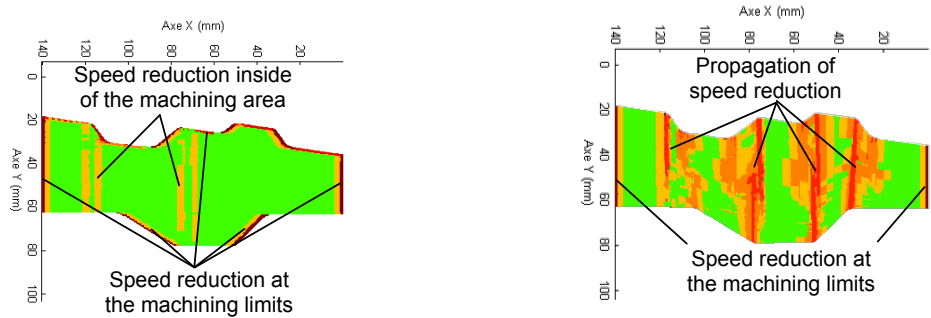

**b. First machining simulation**

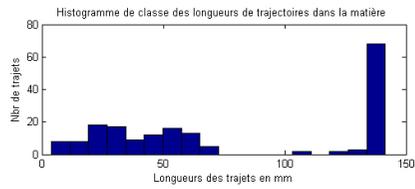 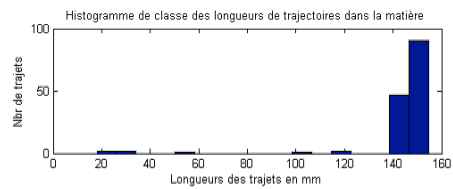

**c. Classification of tool path lengths**

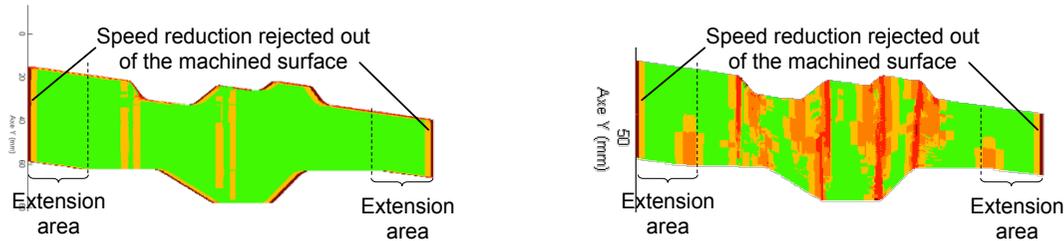

**d. Machining simulation after the extension of tool paths**

**Fig. 7 Influence of topology on speed reduction**

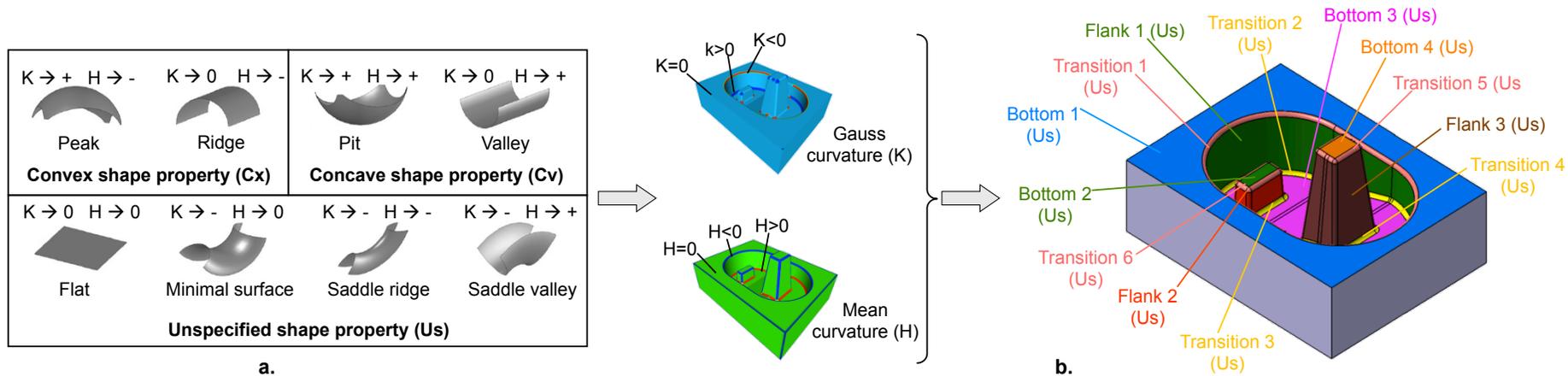

Fig. 8 Principle of topological description

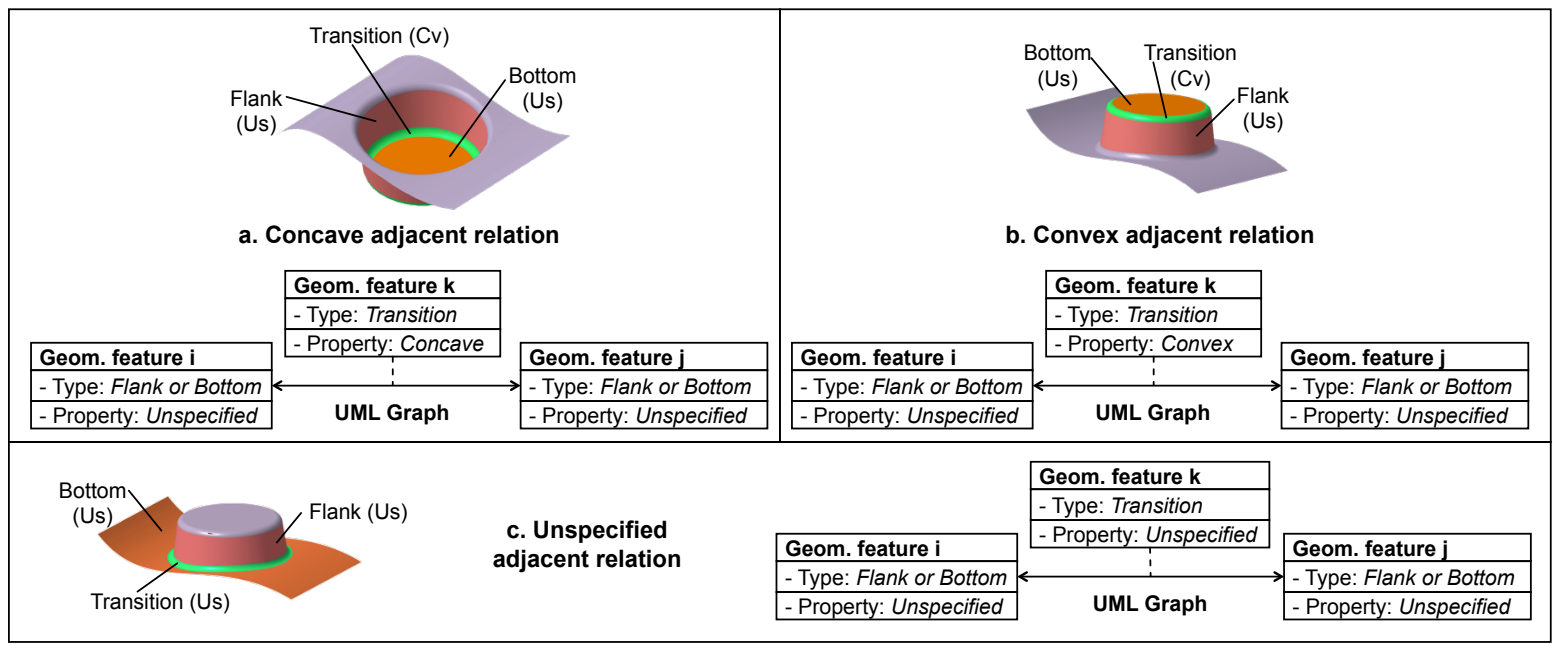

Fig. 9 Adjacent topological relations

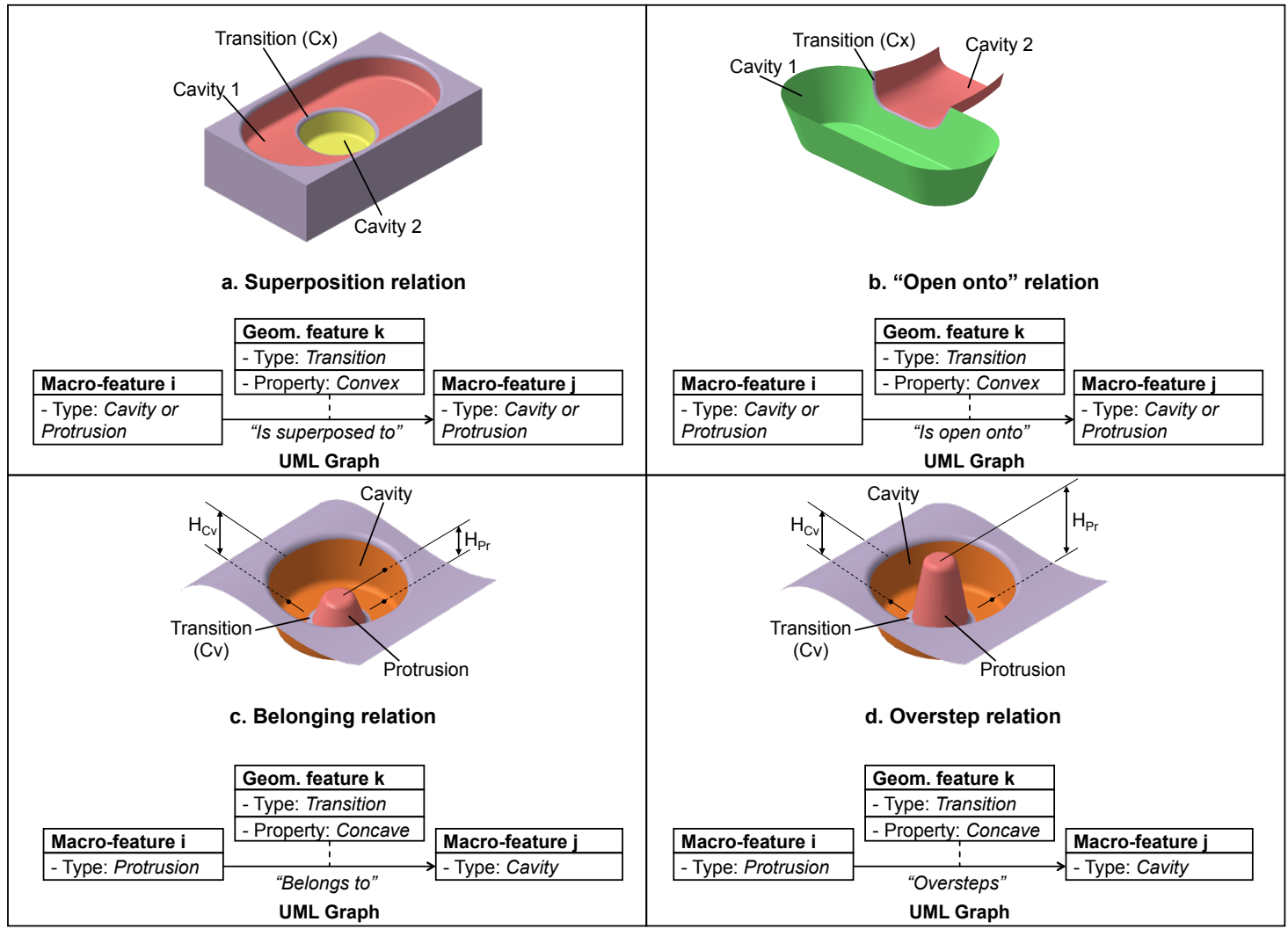

**Fig. 10** Macro-features topological relations

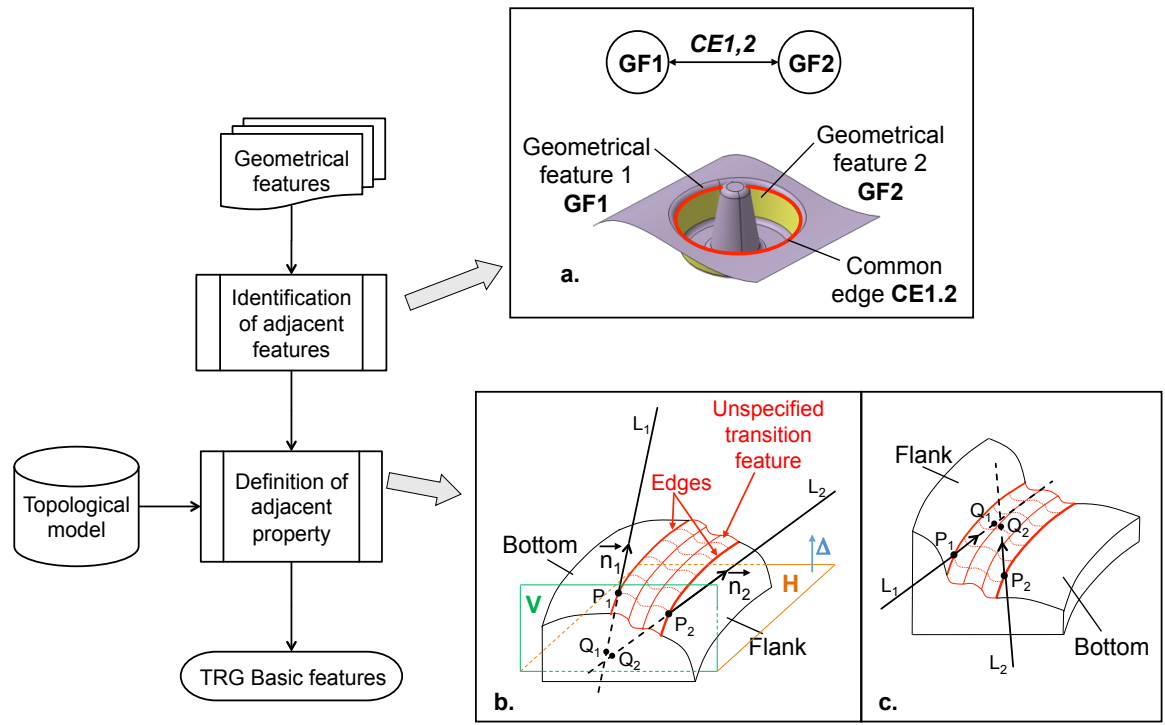

**Fig. 11 Primary graph of topological relations**

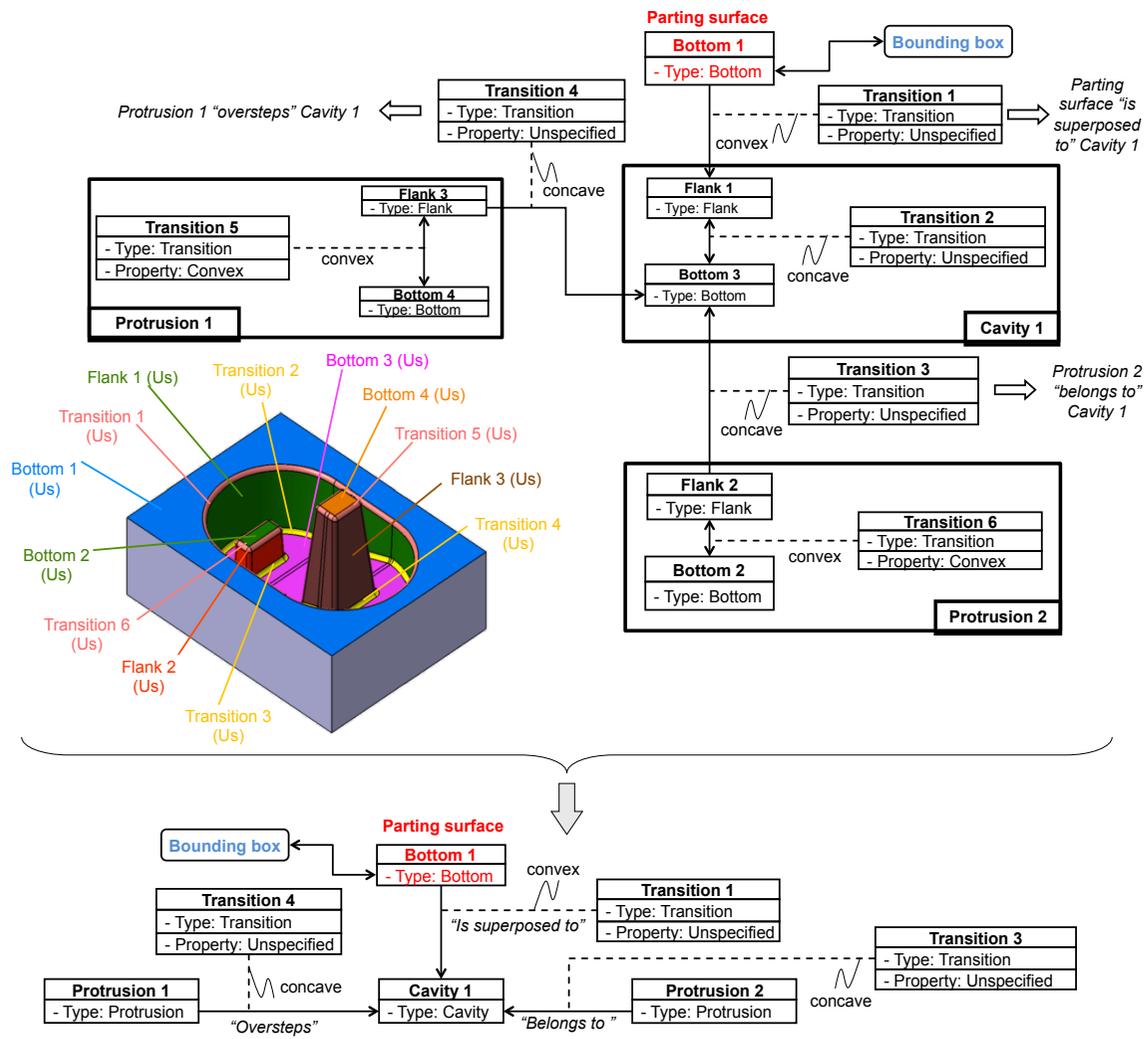

**Fig. 12 Examples of macro-features identification**

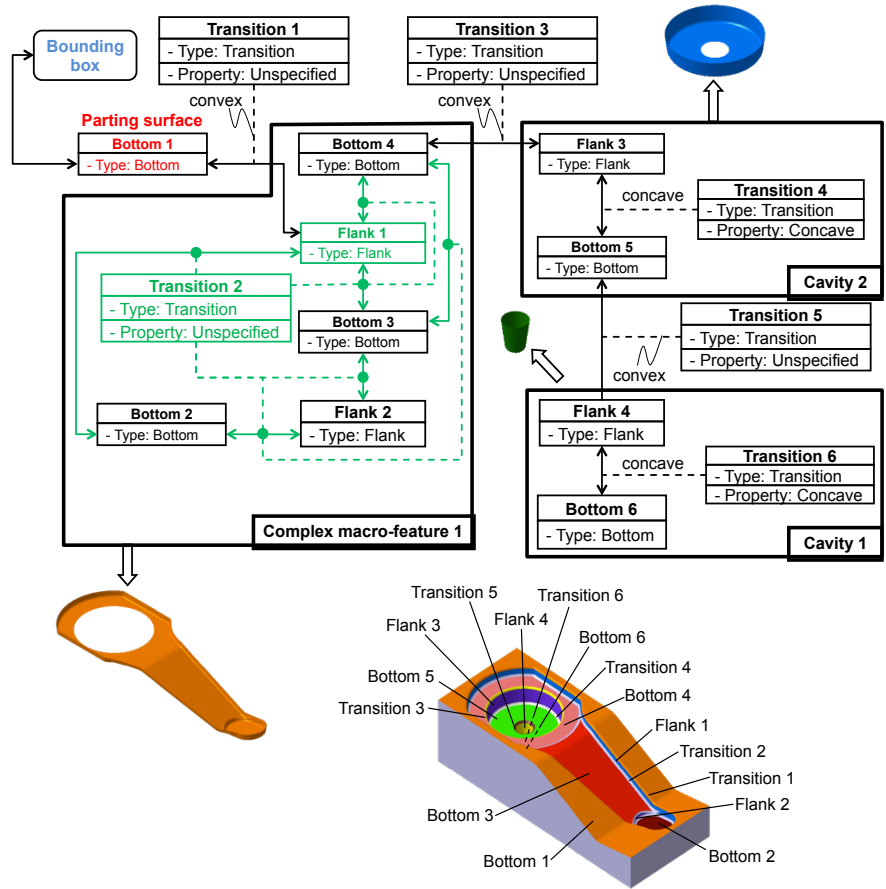

**Fig. 13 Identification of simple and complex macro-features**

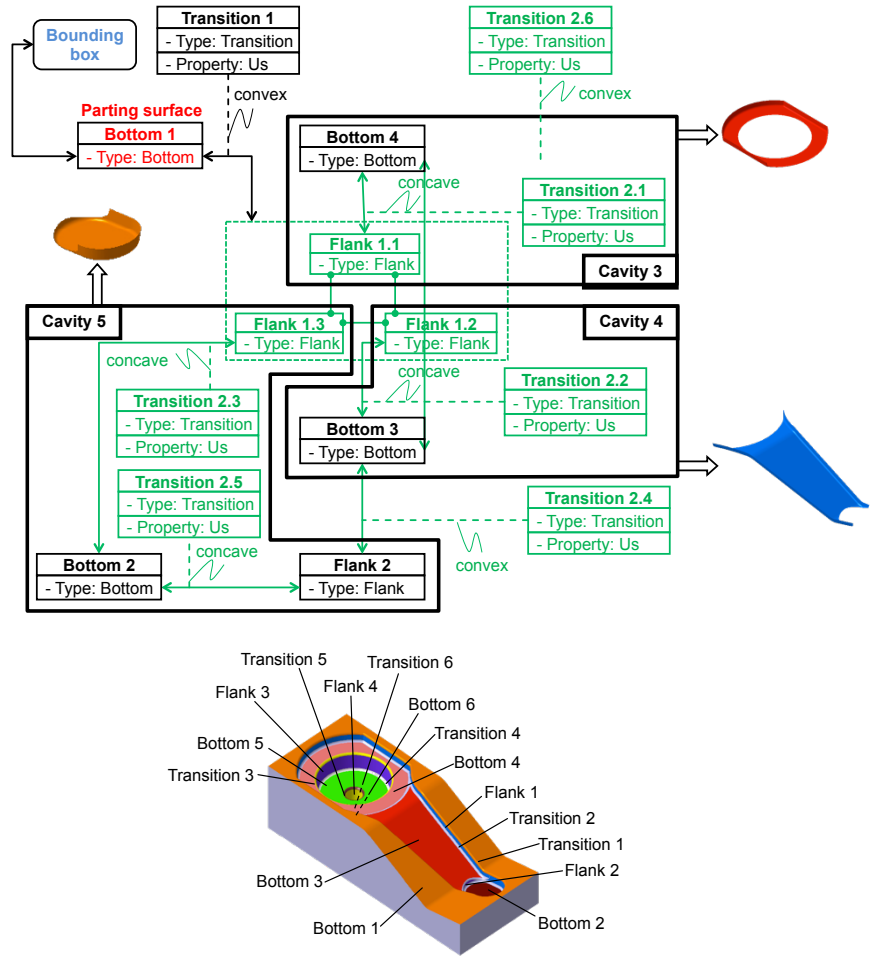

Fig. 14 Decomposition of the complex macro-feature

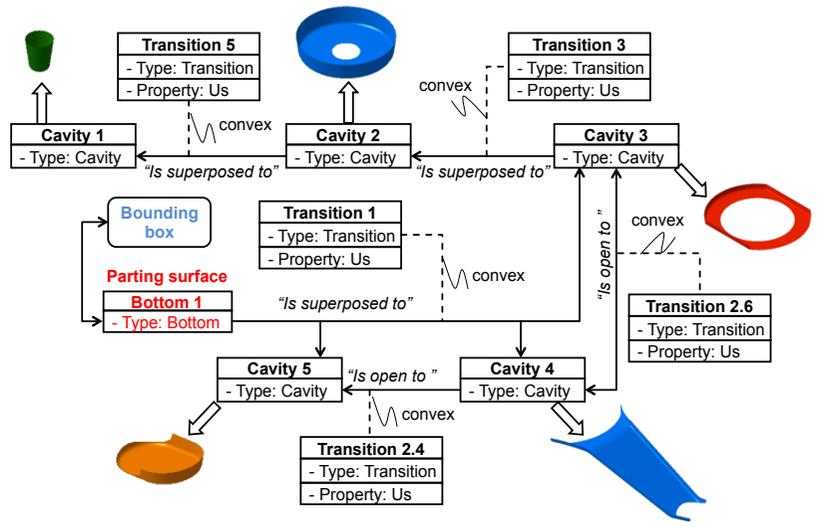

**Fig. 15 Final topological graph of the forging die**